\documentclass[aps,prl,twocolumn,showpacs,amsmath,amssymb]{revtex4}
\usepackage{graphics}

\begin{document}

\title{Comment on ``Berry Phase in a Composite System"}
\author{Li Han}
\author{Yong Guo}\email{guoy66@tsinghua.edu.cn}
\affiliation{Department of Physics, Tsinghua University, Beijing
100084, People's Republic of China}  \maketitle

We show in this comment that in addition to the sign error in
coefficient $d_j$ acknowledged by Yi et al. in their recent erratum
\cite{erratum} for Ref. \cite{Yi}, the definition of subsystem Berry
phase $\gamma=\sum\limits_jp_j\gamma_j$,
$\gamma_j=\mathrm{i}\int_0^T{\mathrm{d}t\langle E_j(t)\vert
\dot{E}_j(t)\rangle}$ proposed as Eq. (7) in Ref. \cite{Yi} is
ambiguous and not gauge invariant.

The fact is that each $\gamma_j$ has a $2\pi$ uncertainty depending
on phase convention of the eigenstate $\vert E_j(t)\rangle$, which
makes the weighted sum $\sum\limits_jp_j\gamma_j$ to be multi-valued
modulo $2\pi$. To see this $2\pi$ uncertainty, we multiply the
eigenstate $\vert E_j(t)\rangle$ with a time-dependent phase factor
$\mathrm{e}^{-\mathrm{i}\varphi_j(t)}$, where $\varphi_j(t)$ is an
arbitrary smooth function with condition
$\varphi_j(T)=\varphi_j(0)+2\pi m_j$ ($m_j$ is an arbitrary integer)
to ensure that
$\vert\tilde{E}_j(t)\rangle=\mathrm{e}^{-\mathrm{i}\varphi_j(t)}\vert
E_j(t)\rangle$ is cyclic too. In this new gauge, $\gamma_j$
transforms as
$\tilde{\gamma}_j=\mathrm{i}\int_0^T{\mathrm{d}t\langle
\tilde{E}_j(t)\vert\dot{\tilde{E}}_j(t)\rangle}=\gamma_j+2\pi m_j$.
The subsystem Berry phase $\gamma$ then transforms as
$\tilde{\gamma}=\sum\limits_jp_j\tilde{\gamma_j}=\gamma+2\pi\sum\limits_jp_jm_j$.
As a result, the subsystem Berry phase is not invariant modulo
$2\pi$ under a gauge transformation when the Schmidt's coefficient
$p_j$ satisfies $0<p_j<1$, which is the case when the composite
system is entangled. The definition of subsystem Berry phase in Ref.
\cite{Yi} is therefore ambiguous and not well-defined.

To overcome this ambiguity, we propose a proper definition of
subsystem Berry phase as
$\gamma=\arg(\sum\limits_jp_j\mathrm{e}^{\mathrm{i}\gamma_j})$,
where it is constructed from a weighted sum of individual phase
factors rather than phases. The phase factor
$\mathrm{e}^{\mathrm{i}\gamma_j}$ eliminates the $2\pi$ uncertainty
in each $\gamma_j$ and makes the new definition to be manifestly
gauge invariant. This definition coincides with the geometric phase
defined in Ref. \cite{Singh} for a mixed state. Based on this
definition, the relation found from Eq. (9) in Ref. \cite{Yi} that
the Berry phases of the subsystems add up to be that of the
composite system no longer holds generally, except when the system
is unentangled \cite{Sjoqvist}.

\begin{figure}
\includegraphics{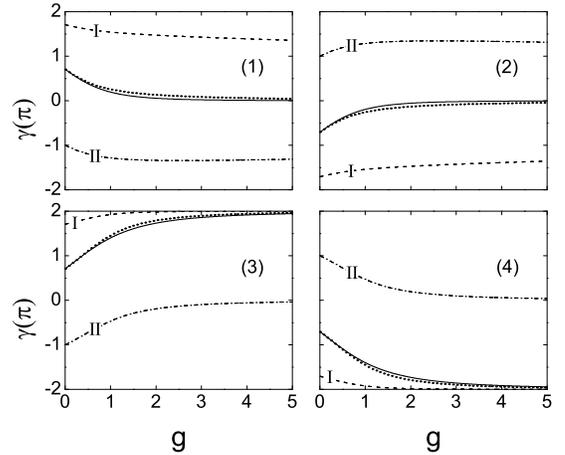}
\caption{\label{Fig.1}Berry phases of the composite system (solid
line), subsystem $\mathrm{I}$ (dashed line), subsystem $\mathrm{II}$
(dash-dotted line), and sum of Berry phases of the subsystems
(dotted line) at $\theta=\frac{\pi}{4}$.}
\end{figure}

We present in Fig. \ref{Fig.1} the Berry phases of the composite
system (denote as $\gamma(m)$, $m=1,2,3,4$) and its subsystems
(denote as $\gamma_{\mathrm{I,II}}(m)$), using the correct form of
coefficient $d_j$ and the proper definition of subsystem Berry
phase. The plots are considerably different from Fig. 6 in Ref.
\cite{Yi} in that (1) $\gamma(1)=-\gamma(2)$, $\gamma(3)=-\gamma(4)$
(mod $2\pi$) and (2)
$\gamma(m)\neq\gamma_{\mathrm{I}}(m)+\gamma_{\mathrm{II}}(m)$ when
the coupling constant $g\neq0$.

This work was supported by the National Natural Science Foundation
of China (No. 10474052).

\end{document}